\documentstyle[aps,twocolumn,psfig]{revtex}  
\def\etal{{\it et al.}}

\def\lap{\hbox{${_{\displaystyle<}\atop^{\displaystyle\sim}}$}}
\def\gap{\hbox{${_{\displaystyle>}\atop^{\displaystyle\sim}}$}}

\begin{document}

\title{Pulsar Constraints on Neutron Star Structure and Equation of
State}

\author{Bennett Link}
\address{Montana State University, Department of Physics, Bozeman MT
59717; blink@dante.physics.montana.edu; Also Los Alamos National
Laboratory}
\author{Richard I. Epstein}
\address{Los Alamos National Laboratory, Mail Stop D436, Los Alamos, NM
87545; epstein@lanl.gov}
\author{James M. Lattimer}
\address{Department of Physics and Astronomy, State University of New
York, Stony Brook, NY 11974-3800; lattimer@astro.sunysb.edu}
\date{\today}
\maketitle
\begin{abstract}


With the aim of constraining the structural properties of neutron
stars and the equation of state of dense matter, we study sudden
spin-ups, {\em glitches}, occurring in the Vela pulsar and in six
other pulsars.  We present evidence that glitches represent a
self-regulating instability for which the star prepares over a waiting
time. The angular momentum requirements of glitches in Vela indicate
that $\ge 1.4$\% of the star's moment of inertia drives these
events. If glitches originate in the liquid of the inner crust, Vela's
`radiation radius' $R_\infty$ must exceed $\simeq 12$ km for a mass of
$1.4 M_\odot$. Observational tests of whether other neutron stars obey
this constraint will be possible in the near future.

\end{abstract}

\section{Introduction}

The sudden spin jumps, or {\em glitches}, commonly seen in isolated
neutron stars are thought to represent angular momentum transfer
between the crust and the liquid interior \cite{models}. In this
picture, as a neutron star's crust spins down under magnetic torque,
differential rotation develops between the stellar crust and a portion
of the liquid interior. The more rapidly rotating component then acts
as an angular momentum reservoir which occasionally exerts a spin-up
torque on the crust as a consequence of an instability. The Vela
pulsar, one of the most active glitching pulsars, typically undergoes
fractional changes in rotation rate of $\sim 10^{-6}$ every three
years on average\cite{vela}. With the Vela pulsar having exhibited 13
glitches, meaningful study of the statistical properties of these
events is now possible.

In this Letter we study the time distribution of Vela's glitches and
determine the average angular momentum transfer rate in Vela and in
six other pulsars. We present evidence that glitches in Vela represent
a self-regulating instability for which the star prepares over a
waiting interval. We obtain a lower limit on the fraction of the
star's liquid interior responsible for glitches.  Assuming that
glitches are driven by the liquid residing in the inner crust, as in
most glitch models, we show that Vela's `radiation radius' is
$R_\infty\gap 12$ km for a mass of $1.4 M_\odot$.  Future measurements
of neutron star radii will check the universality of this constraint
and hence test our understanding of neutron star structure and the
origin of glitches.

\section{Regularity of Angular Momentum Transfer}

A glitch of magnitude $\Delta\Omega_i$ requires angular momentum
\begin{equation}
\Delta J_i = I_c \Delta\Omega_i,
\end{equation}
where $I_c$ is the moment of inertia of the solid crust plus any
portions of the star tightly coupled to it.  Most of the core liquid
is expected to couple tightly to the star's solid component, so that
$I_c$ makes up at least 90\% of the star's total moment of
inertia\cite{core}. Glitches are driven by the portion of the liquid
interior that is differentially rotating with respect to the crust.
The cumulative angular momentum imparted to the crust over time is
\begin{equation}
J(t) = I_c\bar{\Omega}\sum_{i}
 {\Delta\Omega_i\over\bar{\Omega}},
\end{equation}
where $\bar{\Omega}= 70.4$ rad s$^{-1}$ is the average spin rate of
the crust over the period of observations. Fig. 1 shows the cumulative
dimensionless angular momentum, ${ J}(t)/I_c\bar{\Omega}$, over $\sim
30$ years of glitch observations of the Vela pulsar, with a linear
least-squares fit.  The average rate of angular momentum transfer
associated with glitches is $I_c\bar{\Omega}A$, where $A$ is the
slope of the straight line in Fig. 1:
\begin{equation}
A =
(6.44\pm 0.19)\times 10^{-7}\,\,{\rm yr}^{-1}.
\label{jdot}
\end{equation}
This rate $A$ is often referred to as the {\it pulsar activity parameter}.

The angular momentum flow is extremely regular; none of Vela's 13
glitches caused the cumulative angular momentum curve to deviate from
the linear fit shown in Fig. 1 by more than 12\%. To assess the
likelihood that the linear trend could have arisen by chance, we
tested the statistical robustness of this result. We generated many
sets of simulated data in which the occurrence times of the glitches
remained as observed, but the magnitudes of the 13 glitches were
randomly shuffled. We compared the observed $\chi^2$ to those for the
deviations of the randomly shuffled data from linear fits. The
$\chi^2$ for the shuffled data was less than that of the real $\chi^2$
in only $\sim 1.4$\% of cases, strongly suggesting that the rate of
angular momentum flow associated with glitches is reasonably constant.

Additionally, the near uniformity of the intervals between the
glitches in Fig. 1 suggests that glitches occur at fairly regular time
intervals. The standard deviation in observed glitch intervals is
$0.53\langle\Delta t\rangle$, where $\langle\Delta t\rangle = 840$ d
is the average glitch time interval. The probability of 13
randomly-spaced (Poisson) events having less than the observed
standard deviation is only $\sim 1$\%.

The data of Fig. 1 indicate that Vela's glitches are not random, but
represent a self-regulating process which gives a
relatively constant flow of angular momentum to the crust with
glitches occurring at fairly regular time intervals.

\section{The Glitch Reservoir's Moment of Inertia}

The average rate of angular momentum transfer in Vela's glitches
constrains the properties of the angular momentum reservoir that
drives the spin jumps. In particular, the frequent occurrence of large
glitches requires that a significant fraction of the interior
superfluid spins at a higher rate than the crust of the star. Between
glitches, the reservoir acquires excess angular momentum as the rest
of the star slows under the magnetic braking torque acting on the crust.
Excess angular momentum accumulates at the maximum rate 
if the reservoir does not spin down between glitches.
Hence, the rate at which the reservoir accumulates angular momentum
capable of driving glitches is limited by
\begin{equation}
\dot{J}_{\rm res} \le I_{\rm res} |\dot{\Omega}|,
\end{equation}
where $\dot{\Omega}$ is the average spin-down rate of the
crust, and $I_{\rm res}$ is the moment of inertia of the angular momentum
reservoir (not necessarily one region of the star).
Equating $\dot{J}_{\rm res}$ to the average rate of
angular momentum transfer to the crust,
$I_c\bar{\Omega}A$, gives the constraint,
\begin{equation}
{I_{\rm res}\over I_c} \ge
{\bar{\Omega}\over\vert\dot{\Omega}\vert} A
\equiv G,
\label{constraint}
\end{equation}
where the {\em coupling parameter} $G$ is the minimum fraction of the
star's moment of inertia that stores angular momentum and imparts it
to the crust in glitches.  Using the observed value of Vela's activity
parameter $A$ and $\bar{\Omega}/\vert\dot{\Omega}\vert=22.6$ Kyr, we
obtain the constraint
\begin{equation} 
{I_{\rm res}\over I_c} \ge G_{\rm Vela}= 1.4\%.
\end{equation}
A similar analysis for six other pulsars yields the results shown in
Fig.  2. An earlier analysis of glitches in Vela gave $I_{\rm
res}/I_c\ge 0.8$\%\cite{lev}. After Vela, the most significant limit
is obtained from PSR 1737-30 which gives $I_{\rm res}/I_c\ge G_{\rm
1737}= 1$\%.

The similarity of $G$ for the five objects of intermediate age
suggests that glitches in all these objects are driven by internal
components with about the same fractional moment of inertia.  In terms
of $G$, the Crab pulsar and PSR 0525+21 appear to be unusual.  It may
be that the Crab's angular momentum reservoir loses its excess
angular momentum between glitches, perhaps through thermal creep of
superfluid vortices (see, e.g.,
\cite{creep}). The value of $G$ for PSR 0525+21 is not well
determined, since only two glitches from this object have been
measured.

\section{Implications for the Dense Matter Equation of State}

The constraint of $I_{\rm res}/I_c\ge 1.4$\% for Vela applies
regardless of where in the star glitches originate. Many glitch
models, however, assume that the internal angular momentum reservoir
is the superfluid that coexists with the inner crust lattice
\cite{models}, where the pinning of superfluid vortex lines sustains a
velocity difference between the superfluid and the crust. Here we
explore the implications of this interpretation.  We begin by
describing how the moment of inertia of the superfluid in the neutron
star crust relates to the nuclear matter equation of state (EOS) and the
observable properties of neutron stars.

Ravenhall \& Pethick \cite{rp}
have shown that, for various equations of state, the total moment of
inertia
$I$ is given by the approximate expression 
\begin{equation}
\left [1 + {2 G I\over R^3 c^2}\right ]  I \simeq {8\pi\over3}\int_0^R
r^4(\rho+P/c^2) e^{\lambda} dr\equiv \tilde{J}
,
\label{I}
\end{equation}
where $\rho$ is the mass-energy density, $P$ is the pressure, and
$e^{\lambda}$ is the local gravitational redshift.  This expression,
which holds in the limit of slow rotation, defines the integral
$\tilde{J}$.  This integral can be evaluated following Lattimer
\& Prakash
\cite{lp} who noted that
$\rho\propto 1-(r/R)^2$ throughout most of the
interior of a neutron star (but {\it not} in the crust), for
all commonly-used equations of state.  With this
approximation, it can be shown
\cite{lp} that
\begin{equation}
\tilde{J}\simeq {2\over 7} MR^2\Lambda\,,
\label{J}
\end{equation}
where $\Lambda\equiv(1-2GM/Rc^2)^{-1}$  and $M$ is the total stellar
mass.

Equation (\ref{I}) can also be used to determine the moment of inertia
of the crust plus liquid component. In the crust $P$ is $<<\rho c^2$,
and the TOV equation is
\begin{equation}
{dP\over dr} \simeq - G M \rho (r) {e^{\lambda}\over r^2}.
\label{tov}
\end{equation}
Using this approximation in eq. (\ref{I}) gives the fraction of the
star's moment of inertia contained in the solid crust (and the neutron
liquid that coexists with it):
\begin{equation}{\Delta I\over
I}\simeq{8\pi\over3\tilde{J}}\int_{R-\Delta R}^R \rho r^4e^\lambda
dr\simeq {8\pi\over3\tilde{J}GM}\int_0^{P_t}r^6dP\,.
\label{deltai}
\end{equation}
Here $\Delta R$ is the radial extent of the crust and $P_t$ is the
pressure at the crust-core interface. A similar approximation is
obtained in Ref. 7 (equation 17); either approximation is adequate for
the estimates we are making here.
In most of the crust, the equation
of state has the approximately polytropic form $P \propto
\rho^{4/3}$, giving\cite{lp}
\begin{equation}\int_0^{P_t}r^6dP\simeq
P_tR^6\left[1+{8P_t\over n_t
m_nc^2}{4.5+(\Lambda-1)^{-1}\over\Lambda-1}\right]^{-1}\,,
\label{integral}
\end{equation}
where $n_t$ is the density at the core-crust transition and $m_n$ is
the neutron mass.  $\Delta I/I$ can thus be expressed as a function of
$M$ and $R$ with an additional dependence upon the EOS arising through 
the values of $P_t$ and $n_t$.  However, $P_t$ is the main EOS
parameter as $n_t$ enters chiefly via a correction term.
In general, the EOS parameter $P_t$ varies over the range
$0.25<P_t<0.65$ MeV fm$^{-3}$ for realistic equations of state
\cite{lp}. Larger values of $P_t$ give larger values for $\Delta
I/I$, as can be seen from eq. [\ref{integral}].
 
Combining of eqs. [\ref{deltai}] and [\ref{integral}] with a lower
limit on $\Delta I$ and an upper limit on $P_t$ gives a lower limit on
the neutron star radius for a given mass. In order to relate our
observational bound on $I_{\rm res}/I_c$ to $\Delta I$, we assume that
the angular momentum reservoir is confined to the neutron superfluid
that coexists with the nuclei of the inner crust. In this case,
$I_{\rm res}\le\Delta I$ and $I_c\ge I-\Delta I$. Our observational
limit on $I_{\rm res}$ then gives $\Delta I/( I-\Delta I)\ge \Delta
I/I_c \ge I_{\rm res}/I_c \ge 0.014$. To obtain a strong lower limit
on the neutron star radius, we take $P_t=0.65$ MeV fm$^{-3}$ and
$n_t=0.075$ fm$^{-3}$. Combining the relations [\ref{deltai}] and
[\ref{integral}], gives the heavy dashed curve in Fig. 3.  This curve
is given approximately by
\begin{equation}
R=3.6+3.9 M/M_\odot\,.
\label{rlimit}
\end{equation}
Stellar models that are compatible with the lower bound on $I_{\rm
res}$ must fall below this line. Smaller $P_t$ reduces the
crustal moment of inertia and gives a more restrictive constraint. For
example, $P_t=0.25$ MeV fm$^{-3}$ moves the constraining contour to
approximately $R=4.7+4.1M/M_\odot$ (thin dashed curve of Fig. 3).

\section{Discussion}

To summarize our conclusions regarding the statistics of Vela's
glitches, we find that angular momentum is imparted to the crust at
regular time intervals at a rate that has remained nearly constant for
$\sim 30$ yr. These data narrowly constrain the {\em coupling
parameter} $G$ which is the minimum fraction of the star's moment of
inertia that is responsible for glitches. For Vela we find $G=0.014$,
indicating that least $1.4$\% of the star's moment of inertia acts as
an angular momentum reservoir for driving the glitches, regardless of
where in the star this angular momentum reservoir is, or how it is
coupled to the crust.  Variation of $G$ by a factor of less than $\sim
3$ for stars in the age group $10^4-10^5$ yr suggests that glitches in
stars in this age group all involve regions of about the same
fractional moment of inertia.

Mass measurements of radio pulsars in binary systems and of neutron star
companions of radio pulsars give neutron star masses consistent with a
very narrow distribution, $M=1.35\pm 0.04 M_\odot$ \cite{masses},
indicated by the pair of horizontal dotted lines in Fig. 3. If Vela's
mass falls in this range, eq. [\ref{rlimit}] constrains $R\gap 8.9$
km, under the assumption that glitches arise in the inner crust
superfluid. The quantity constrained by observations of the stellar
luminosity and spectrum is the `radiation radius'
$R_\infty\equiv\Lambda^{1/2}R =(1-2GM/Rc^2)^{-1/2}R$. If
$M=1.35M_\odot$ for Vela, the above constraint gives $R_\infty\gap
12$ km if glitches arise in the inner crust.  For comparison, we show
in Fig. 3 the mass-radius curves for several representative equations
of state (heavy solid lines). Measurement of $R_\infty\lap 13$ km
would be inconsistent with most equations of state if $M\simeq 1.35
M_\odot$. Stronger constraints could be obtained if improved
calculations of nuclear matter properties indicate $P_t$ significantly
less than 0.65 MeV fm$^{-3}$. For example, for $M\simeq 1.35 M_\odot$,
$R_\infty\gap 13$ km would be required if $P_t=0.25$ MeV fm$^{-3}$. A
measurement of $R_\infty\lap 11$ km would rule out most equations of
state regardless of mass or the angular momentum requirements of
glitches.

A promising candidate for a decisive measurement of a neutron star's
radiation radius is RX J185635-3754, an isolated, non-pulsing neutron
star \cite{walter}. A black body fit to the X-ray spectrum gives
$R_\infty =7.3 (D/120\ {\rm pc})$ km where $D$ is the distance (known
to be less than 120 pc).  However, either a non-uniform surface
temperature or radiative transfer effects in the stellar atmosphere
could raise this estimate significantly \cite{alpw}.  HST observations
planned for this year should determine the star's proper motion and
parallax, and hence, the distance.  Future CHANDRA observations should
yield more detailed spectral data and could establish the composition
of the atmosphere if absorption lines are identified. These distance
and spectral data may establish whether this object's radius is consistent
with an inner crust explanation of neutron star glitches.

We thank P. M. McCulloch for providing us with glitch data for the
Vela pulsar.  This work was performed under the auspices of the U.S.
Department of Energy, and was supported in part by NASA EPSCoR Grant
\#291748, NASA ATP Grants
\# NAG 53688 and \# NAG 52863, by the USDOE grant
DOE/DE-FG02-87ER-40317, and by IGPP at LANL.

\newpage

\vbox{\hspace*{-1.cm}\psfig{figure=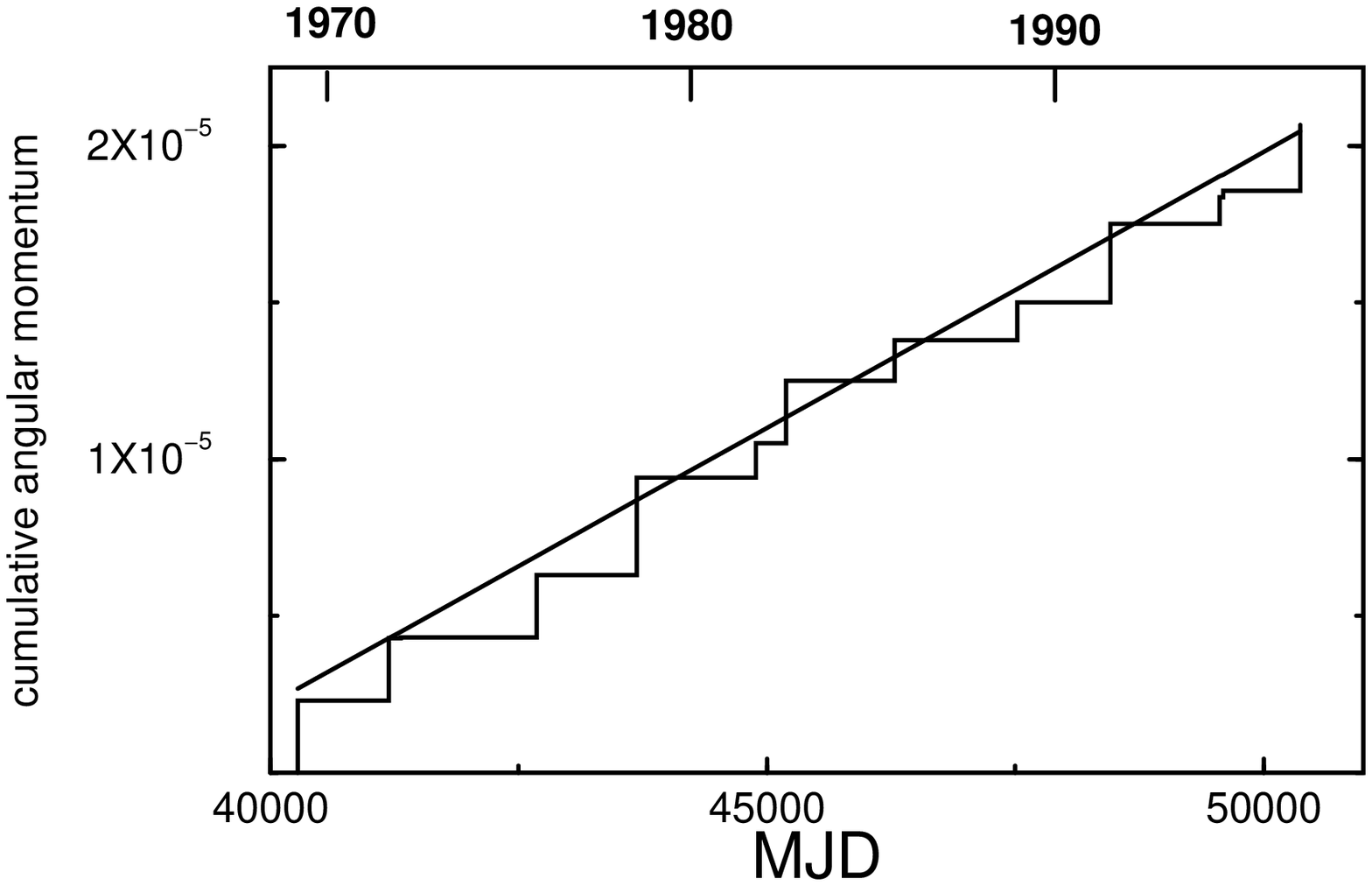,width=4.in}}
\noindent Fig. 1 -- Cumulative dimensionless angular momentum,
$J/I_c\bar{\Omega}$, imparted to the Vela pulsar's crust as a function
of time. The straight line is a least-squares fit.

\vspace*{1.cm}

\vbox{\hspace*{-1.cm}\psfig{figure=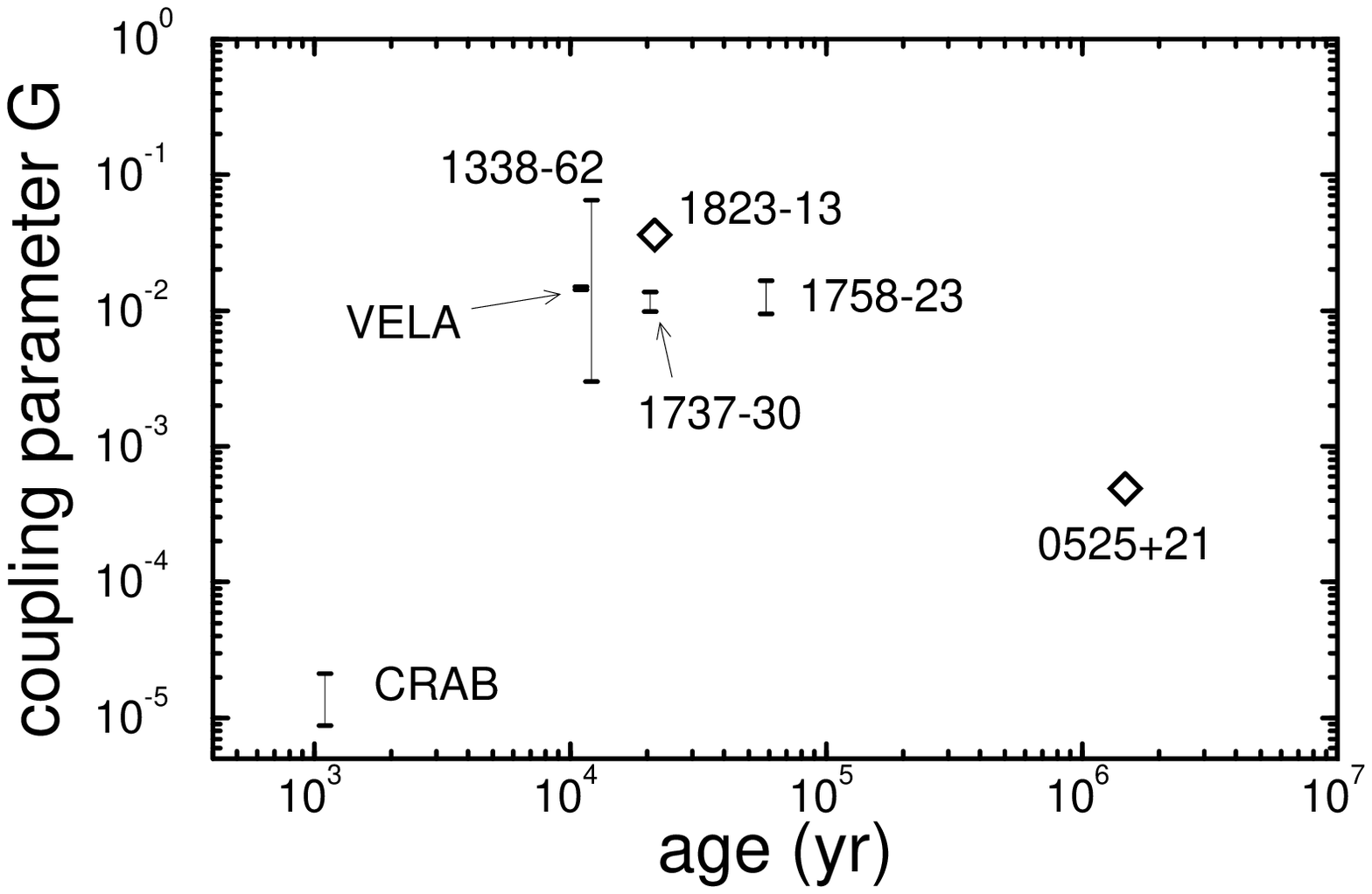,width=4.in}}
\noindent Fig. 2 -- The coupling parameter G.  The strongest constraints are
obtained for Vela and PSR 1737-30, for which 13 and 9 glitches have
been observed, respectively.  Diamonds indicate objects with only two
observed glitches, for which error bars could not be obtained.
References:  0525+21 [14], Crab [15], Vela [2], 
1338-62 [16], 1737-30 [17,18], 1823-13 [18].

\newpage

\vbox{\hspace*{-1.5cm}\psfig{figure=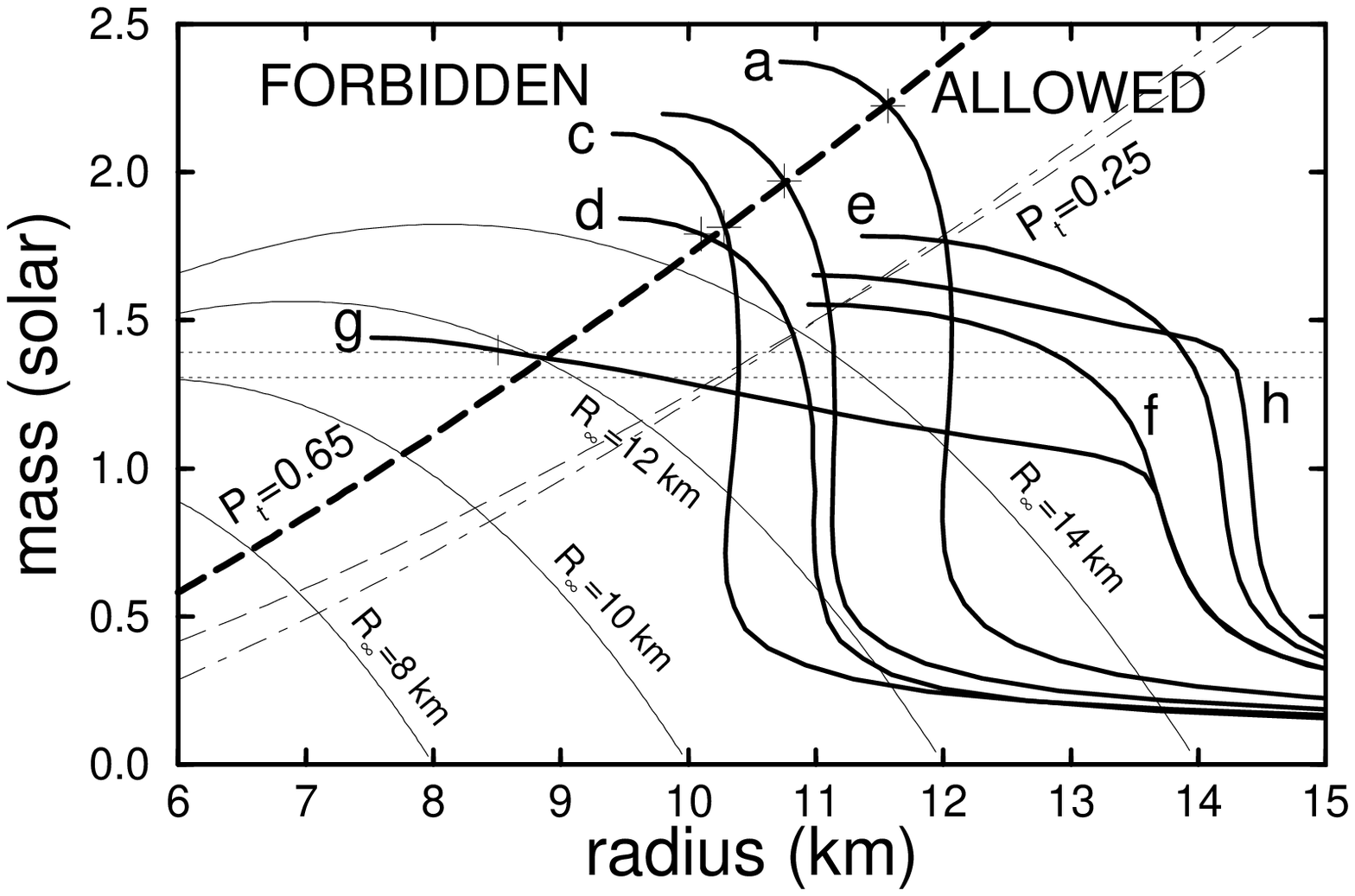,width=4.in}}
\noindent Fig. 3 -- Limits on the Vela pulsar's radius. The heavy dashed curve
delimits allowed masses and radii that are compatible with the glitch
constraint $\Delta I/(I-\Delta I)\ge 1.4$\% for $P_t=0.65$ MeV
fm$^{-3}$. The thin dashed curve corresponds to $P_t=0.25$ MeV
fm$^{-3}$ and gives a more stringent though less conservative
constraint. The dot-dashed curve corresponds to $\Delta I/(I-\Delta
I)\ge 2.8$\% and $P_t=0.65$ MeV. The horizontal dashed lines indicate
the mass limits for the survey of 26 radio pulsars of Ref. 19. Also
displayed are mass-radius relations for the equations of state of
Akmal \& Pandharipande [10] (curves a and b), Wiringa, Fiks \&
Fabrocini [11] (curves c and d), M\"uller \& Serot [12] (curves e and
f) and the kaon EOS of Glendenning \& Schaffner-Bielich [9] (curves g and
h). The crosses indicate where a given EOS has $\Delta I/(I-\Delta
I)=1.4$\% (with $P_t=0.65$ MeV fm$^{-3}$). Curves without crosses have
$\Delta I/(I-\Delta I)>1.4$\% for all stable $R$. Thin curves are
contours of constant radiation radius $R_\infty$. 


\begin{references}

\def\nature{{\rm Nature}}
\def\nucphys{{\rm Nuc.Phys}}
\def\nucphysa{{\rm Nuc. Phys. A}}
\def\physletb{{\rm Phys. Lett. B}}
\def\physrevc{{\rm Phys. Rev. C}}
\def\prd{{\rm Phys. Rev. D}}
\def\sovphysjetp{{\rm Soviet~Phys.~JETP}}
\def\ptpl{{\rm Progr.Theor.Phys.Lett}}
\def\ptps{{\rm Prog. Theor. Phys. Suppl.}}
\def\ptp{{\rm Prog. Theor. Phys.}}

\bibitem{models}
P. W. Anderson and N. Itoh, \nature, {\bf 256}, 25 (1975);
M. Ruderman, {\bf 203}, 213 (1976);
D. Pines and M. A. Alpar, \nature, {\bf 316}, 27 (1985).

\bibitem{vela} J. M. Cordes,
G. S. Downs and J. Krause-Polstorff, \apj, {\bf 330}, 847 (1988);
P. M. McCulloch, \etal\ Aust. J. Phys., {\bf 40}, 725 (1987);
C. Flanagan, IAU Circ. No. 4695 (1989);
C. Flanagan, IAU Circ. No. 5311 (1991).

\bibitem{core}
M. A. Alpar, S. A. Langer and J. A. Sauls, \apj, {\bf 282}, 533
(1984); M. Abney, R. I. Epstein and A. Olinto, \apj, {\bf 466}, L91
(1996).

\bibitem{lev}
B. Link, R. I Epstein and K. A. Van Riper, \nature, {\bf 359}, 616
(1992). 

\bibitem{lyne} A. G. Lyne, {\sl Lives of the
Neutron Stars}, p. 167. Ed: M. A. Alpar (Kluwer, 1995).

\bibitem{creep}
M. A. Alpar, P. W. Anderson and D. Pines, \apj, {\bf 276}, 325 (1984);
B. Link, R. I. Epstein, G. Baym, \apj, {\bf 403}, 285 (1993);
H. F. Chau and K. S. Cheng, K. S., \prb, {\bf 47}, 2707 (1993).

\bibitem{rp}
D. G. Ravenhall and C. J. Pethick, \apj, {\bf 424}, 846 (1994).

\bibitem{lp}
J. M. Lattimer and M. Prakash, in preparation (1999).

\bibitem{gs} N. K. Glendenning and J. Schaffner-Bielich,
Phys. Rev. C, {\bf 60}, 25803 (1999).

\bibitem{ap} A. Akmal and V. R. Pandharipande, Phys. Rev. C, {\bf 56},
2261 (1997).

\bibitem{wff} R. B. Wiringa, V. Fiks and A. Fabrocini, Phys. Rev., C,
{\bf 38}, 1010 (1988).

\bibitem{ms} H. M\"uller and B. D. Serot, Nucl. Phys., {\bf 606}, 508
(1996).

\bibitem{walter} F. Walter, S. Wolk and R. Neuhauser, \nature, {\bf
379}, 233 (1996).

\bibitem{downs82} G. S. Downs, \apj, {\bf 257}, L67 (1982).

\bibitem{crab}
P. E. Boynton \etal\, \apj, {\bf 175}, 217 (1972);
E. Lohsen, \nature, {\bf 258}, 688 (1975);
A. G. Lyne and R. S. Pritchard, MNRAS, {\bf 229}, 223 (1987);
A. G. Lyne, R. S. Pritchard and F. G. Smith, \nature, {\bf 359}, 706
(1992).

\bibitem{kaspi92} V. M. Kaspi, R. N. Manchester,
S. Johnston, A. G. Lyne and N. D'Amico, \apj, {\bf 399}, L155 (1992).

\bibitem{ml}
A. G. McKenna and A. G. Lyne, \nature, {\bf 343}, 349 (1990).

\bibitem{sl}
S. L. Shemar and A. G. Lyne, MNRAS, {\bf 282}, 677 (1996).

\bibitem{masses}
S. E. Thorsett and D. Chakrabarty, \apj, {\bf 512}, 288. 

\bibitem{alpw}
P. An, J. M. Lattimer, M. Prakash, and Walter, F. in preparation
(1999). 

\end{references}
\end{document}